\documentclass[a4paper,10pt]{article}
\usepackage[utf8]{inputenc}
\usepackage{xspace}			   
\usepackage[english]{babel}
\usepackage[T1]{fontenc}
\usepackage{lmodern}
\usepackage{authblk}

\usepackage[margin=25mm,includeheadfoot,bindingoffset=10mm]{geometry}
\usepackage{setspace}
\usepackage{ragged2e}
\usepackage[all]{xy}
\usepackage{graphicx} 
\usepackage{color}
\usepackage[svgnames]{xcolor}

\usepackage{pdfpages}
\usepackage{amsmath, mathtools}
\usepackage{amssymb,amsfonts,dsfont}
\usepackage{mathrsfs}
\usepackage{amsthm}
\usepackage{epigraph}
\usepackage{bm}				
\usepackage{bbm,upgreek}

\usepackage{hyperref} 
\hypersetup{linktocpage,colorlinks,citecolor={blue},pdfdisplaydoctitle=true,pdfpagemode=UseOutlines,bookmarksnumbered=true}
\usepackage[sort&compress,numbers]{natbib}
\usepackage{url}
\usepackage{doi}

\newcommand{\bra}[1]{\langle #1 |} 
\newcommand{\ket}[1]{| #1 \rangle } 

\definecolor{cbl}{rgb}{0,0,1}

\definecolor{crd}{rgb}{1,0,0}
 
\newcommand{\upd}{\mathrm{d}}

\newcommand{\xb}{\mathbf{x}}
\newcommand{\yb}{\mathbf{y}}

\newcommand{\ie}[0]{\textit{i.e.} }
\newcommand{\eg}[0]{\textit{e.g.} }

\title{Binding quantum matter and space-time, without romanticism}

\author{\vskip0.2cm Antoine Tilloy\footnote{\url{antoine.tilloy@mpq.mpg.de}}}

\affil{\vskip-0.2cm \it \small Max-Planck-Institut f\"ur Quantenoptik, Hans-Kopfermann-Stra{\ss}e 1, 85748 Garching, Germany \normalsize}

\date{\vskip-0.6cm\small (\today) \normalsize}

\begin{document}

\maketitle
\begin{abstract}
    Understanding the emergence of a tangible 4-dimensional space-time from a quantum theory of gravity promises to be a tremendously difficult task. This article makes the case that this task may not have to be carried. Space-time as we know it may be fundamental to begin with. I recall the common arguments against this possibility and review a class of recently discovered models bypassing the most serious objection. The generic solution of the measurement problem that is tied to semiclassical gravity as well as the difficulty of the alternative make it a reasonable default option in the absence of decisive experimental evidence. 
\end{abstract}

\section{Introduction}
Quantum gravity may profoundly modify the way we understand space-time: the geometric structures inherited from General Relativity may have to be discarded, causality may become available only as an emergent concept, and all the notions we are used to may become irreducibly blurry. All our certitudes may have to go down the drain. Such a revolution would be exhilarating for theoretical physicists and philosophers, offering them the all too rare opportunity to rebuild physics from the ground up. For this reason, it seems that the prudent alternative has been insufficiently discussed. Quantum gravity may eventually change nothing at all.

If quantum matter and a tangible space-time can be coupled in a consistent way, and if this theory is chosen by Nature, then the revolution promised by quantum gravity will simply not happen. This option is arguably much less romantic, but this in itself does not make it less plausible. To paraphrase J. S. Bell, we should not let ourselves be fooled by our natural taste for romanticism \cite{bell1992}. Constructions yielding counterintuitive predictions should be favored only when they become inevitable; grandiose claims should be allowed in physics only when all the boring options have been fully exhausted\footnote{The ``boredom'' razor (consisting in maximizing the boring character of a theory) is likely a better guide than an Ockham's razor based solely on beauty (maximizing beauty or mathematical simplicity).}. As we shall argue, this is far from being the case in the gravitational context. The ``quantization'' route and its numerous conceptual subtleties has been favored on the basis of weak theoretical evidence and in the complete absence of experimental input. Further, after forty years of intense theoretical efforts, progress in this direction has been scarce. This makes it pressing to reconsider the ontologically boring option of a non-quantized space-time, perhaps prematurely discarded.

The objective of this article is to show that a fundamentally semiclassical theory of gravity, with a traditional notion of space-time, should not be a priori ruled out. I am not the first to discuss this option (see \eg \cite{carlip2001,huggett2001,mattingly2005,wuthrich2005}), but recent theoretical results allow to counter historical rebuttals in a constructive way. Indeed, instead of pointing out loopholes or limitations in historical pseudotheorems, I will present a way to construct consistent models of (Newtonian) semiclassical gravity, with a transparent empirical content, thus evading any argument that could possibly exist against them.

As we shall see, there does exist a number of philosophical arguments in favor of the quantization of gravity. However, most of them are weak: either purely aesthetic (``everything should be quantized'') or showing blatant wishful thinking (``a quantum space-time will fix the problems we have with other theories''). Yet, I will identify one argument that would be strong if its premise were correct. It states that semiclassical theories are necessarily inconsistent; constructing hybrid theories coupling quantum and classical variables would inevitably yield paradoxes. 
In a nutshell, this argument rests on a dangerous straw man: taking flawed mean-field semiclassical approaches as representatives of all hybrid quantum-classical theories. Actually, there exists (at least) one perfectly consistent way to couple classical and quantum variables using the machinery (but not the interpretation) of orthodox measurement and feedback. This approach can be instantiated in several ways in the non-relativistic context where it yields consistent candidates for semiclassical theories of Newtonian gravity.

I will first recall the weakest objections to semiclassical gravity before introducing the standard mean field approach and explaining its genuine shortcomings. Then, I will discuss a sound alternative formally grounded on feedback.  Finally, I will comment on the implications of this semiclassical toy model and argue, as advertised, that quantum gravity (understood in the broad sense of a consistent theory of quantum matter and gravity) need not have any radical metaphysical implications for the fundamental nature of spacetime.

\section{The shaky case for quantizing everything}\label{sec:2}
Before going to the heart of the matter and the construction of consistent semiclassical theories of gravity, I have to discuss the common objections raised against this endeavour. 

The first strand of argument rests on hope. Quantizing space-time degrees of freedom may help smooth out the nasty behavior of our current continuum theories. Quantized gravity is optimistically expected to (1) regularize the UV divergences of Quantum Field Theory (2) tame the singularities of General Relativity.

The first point is a reference to the Landau pole \cite{landau1955} in the electroweak sector of the Standard Model. Even though quantum electrodynamics is perturbatively renormalizable, renormalization group heuristics and numerical computations suggest that the theory is trivial in the absence of a short distance cut-off \cite{gies2004}. The quantization of gravity is supposed to provide the granularity of space-time that would save electromagnetism from being only effective. The second point is a reference to the singularities of the metric at the center of Black Holes (\eg with the Schwartzchild metric) or at the Big-Bang in cosmological scenarios (\eg in the Friedmann-Lema\^itre-Robertson-Walker metric). Quantizing gravity might miraculously solve these two kinds of problems (for space-time singularities, the situation is actually far from clear \cite{falciano2015,struyve2017lqg}). Yet in both cases, appealing to quantum gravity in these contexts is dangerous wishful thinking: we are just pushing under the carpet of yet unknown theories the difficulties we have with existing ones. Before the advent of quantum electrodynamics, quantum mechanics was already expected to cure the UV divergences of Maxwell's equations with point charges. No such thing eventually happened and the situation arguably got worse\footnote{Physicists behave a bit like snake oil salesmen. They promise ``quantization'' will do for gravity what it did not achieve for electrodynamics swearing to get the (still ill-defined) quantum electrodynamics cured in the process. How will we solve anything when there is nothing left to quantize?}.  In any case, that quantizing gravity might solve some existing problems does not mean other solutions may not exist. That quantized gravity might be mathematically convenient does not make it a necessity.

The second strand of arguments is aesthetic. Gravity should be quantized because everything should be quantized. Quantum theory should not be seen as a mere theory of matter but as a meta-theory \cite{mattingly2005}, as a set of principles (\eg Heisenberg uncertainty, ``complementarity'', canonical quantization) to be applied to all ``classical'' models. In this respect, gravity should be ``just like the other forces''. And, so the argument goes, History proved that the people who tried to keep the electromagnetic field classical were wrong. But is gravity really like the other forces? Could the tremendous technical and conceptual difficulties encountered when attempting to quantize gravity not be a sign (among others) of its intrinsic difference? Gravity, after all, is not just a spin 2 Gauge field \cite{carlip2001}. Aesthetic arguments, it would seem, can as well be used to argue against the quantization of gravity. And even if quantized gravity were to be superior on aesthetic grounds, would this pay for all the romanticism that comes with it? In my opinion, relying on such a highly subjective and questionable meta theoretical notion of beauty at any cost is simply irrational. Quantum mechanics could very well just be simply a dynamical theory of matter

The last strand of arguments is mathematical and seems to have the sharpness of a theorem. Semiclassical theories, mixing a quantum sector (here matter) and classical one (here space-time), would inherently be inconsistent (in a sense to be later defined). If this is indeed the case, then this argument dwarfs all the previously mentioned unconvincing incantations. Fortunately, looking at it more closely, we shall see that it applies only to a specific way of constructing semiclassical theories. It is a straw-man argument: picking the most naive and ill-behaved approach as a representative for the whole. Semiclassical theories constructed with mean field coupling are inconsistent with some features of quantum theory (like the Born rule), but the way is open for other approaches.

In the end, the case for the quantization of gravity is supported by a single \emph{real} argument: naive attempts at constructing hybrid semiclassical theories have failed. Hence, if we want to know whether or not it is possible to hold on to space-time as we know it, we have to understand the problem with existing hybrid theories and see if and how it can be fixed.

\section{Naive semiclassical gravity}\label{sec:3}

\subsection{General problem}

\begin{figure}
    \centering
    \includegraphics[width=0.6\textwidth]{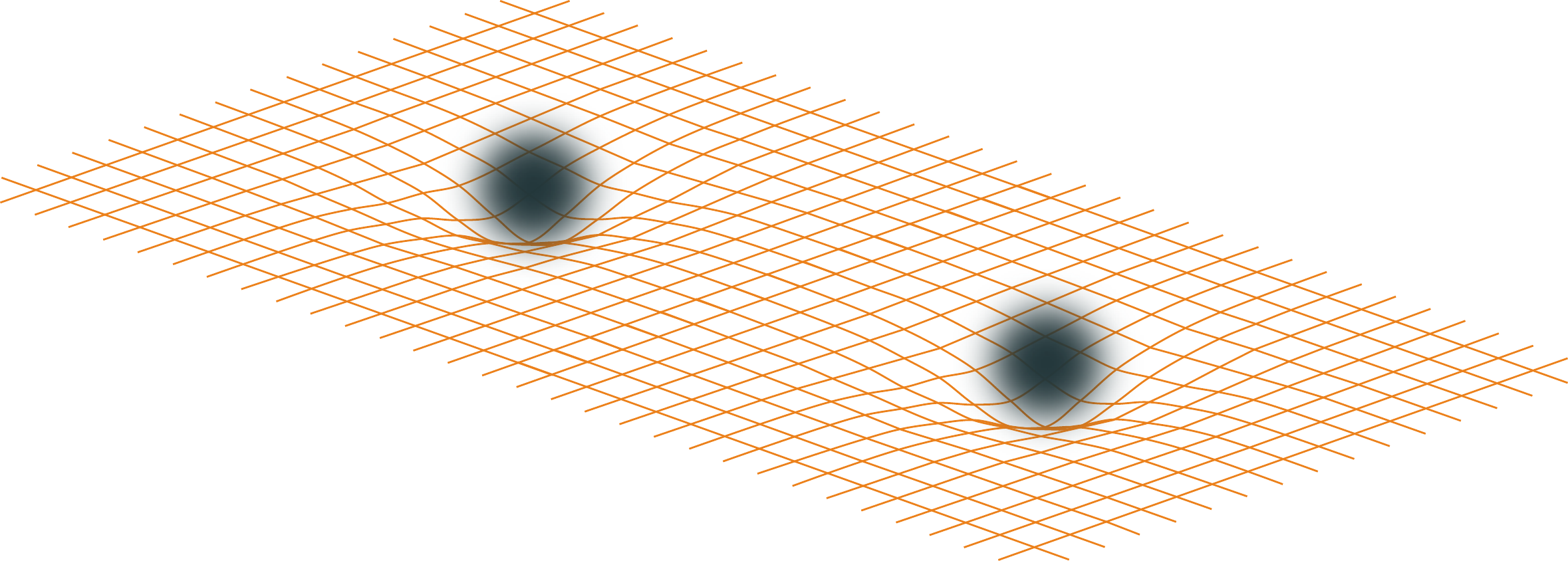}
    \caption{In semiclassical gravity, the \emph{easy problem} is to understand how quantum matter (the fuzzy balls) moves in a curved spacetime (the mesh). The \emph{hard problem} is to understand how spacetime is curved by quantum matter. The naive option, which consists in taking $|\psi|^2(x,t)$ as source (or more precisely, its relativistic, many-particle equivalent) yields insurmountable problems.}
    \label{fig:semiclassical}
\end{figure}

In a semiclassical theory of gravity, a classical space-time and quantum matter should cohabit at the deepest level. Hence, to construct such a theory, one needs to understand (see also Fig. \ref{fig:semiclassical}):
\begin{enumerate}
    \item how quantum matter moves in a curved background,
    \item how space-time is curved by quantum matter.
\end{enumerate}
The first point is uncontroversial and expected to be solved fully by quantum field theory in curved space-time (QFTCST) \cite{wald1994}. In a way, the latter is a very elaborate version of the quantum theory of non-relativistic particles in an external electromagnetic field. This is not to say that there are no technical difficulties or open questions, but there is little doubt about the way to proceed to compute predictions (at least perturbatively). Further, at least in some limits, the theory has been tested experimentally \cite{colella1975,schlippert2014}. The second point, understanding how quantum matter could consistently ``source'' curvature, is the true open problem and the one we will need to address. Indeed, the classical Einstein equation reads:
\begin{equation}\label{eq:einstein}
R_{\mu\nu} -\frac{1}{2} R g_{\mu \nu} =8\pi G T_{\mu\nu},
\end{equation}
where $T_{\mu\nu}$ is the (classical) stress energy tensor. For quantum matter, it is not clear what object to put as a source. Let us first consider the historical and infamous ``mean-field'' option.

\subsection{The Schr\"odinger-Newton equation}

The historical proposal, due to M\o ller and Rosenfeld \cite{moller1962,rosenfeld1963}, is to create a classical stress energy tensor with the help of a quantum expectation value, \ie to posit that:
\begin{equation}\label{eq:mr}
R_{\mu\nu} -\frac{1}{2} R g_{\mu \nu} =8\pi G \, \langle\hat{T}_{\mu\nu}\rangle.
\end{equation}
Let us notice that there are important subtleties involved and giving a precise meaning to this equation in a quantum field theory context is not trivial. In particular, $\langle\hat{T}_{\mu\nu}\rangle$ has to be suitably renormalized to remove the infinite contribution from the vacuum energy \cite{wald1978,wald1994}. However, I will not be concerned by such technicalities and I will consider directly the non-relativistic limit of the M\o ller-Rosenfeld prescription which is mathematically unproblematic. In the non-relativistic limit, Einstein's equation becomes the Poisson equation for a scalar gravitational field and the stress energy tensor reduces to the mass density:
\begin{equation}\label{eq:poisson}
\nabla^2 \Phi(\xb,t)=4\pi G\, \bra{\psi_t} \hat{M}(\xb) \ket{\psi_t},
\end{equation}
where $\Phi(\xb,t)$ is the gravitational field, $\ket{\psi_t}$ is the quantum state of matter and $\hat{M}(\xb)$ is the mass density operator\footnote{For $n$ species of particles of mass $m_k$, $1\leq k\leq n$, the mass density operator is conveniently written in second quantized notation:
\begin{equation}
\hat{M}(\xb)=\sum_{k=1}^n m_k a_k^\dagger(\xb) a_k(\xb),
\end{equation} 
where $a^\dagger_k(\xb)$ and $a_k(\xb)$ are the creation and annihilation operators in position $\xb$ for the particle species $k$.}. As I mentioned earlier, matter then evolves with the classical gravitational field $\Phi_t$ (this latter part is uncontroversial):
\begin{equation}
\frac{\upd}{\upd t} \ket{\psi} = - \frac{i}{\hbar}\left( H_0 + \int \upd \xb\,\Phi(\xb,t) \hat{M}(\xb) \right)\ket{\psi_t},
\end{equation}
where $H_0$ contains the free part Hamiltonian and other non-gravitational interactions.
Inverting the Poisson equation \eqref{eq:poisson} then yields the celebrated Schr\"odinger-Newton equation \cite{diosi1984,bahrami2014}:
\begin{equation}\label{eq:SN}
\frac{\upd}{\upd t} \ket{\psi_t} =-\frac{i}{\hbar}H_0\ket{\psi_t} + i \, \frac{G}{\hbar} \int \upd \xb\,\upd \yb\,\frac{\bra{\psi_t} \hat{M}(\xb)\ket{\psi_t}\, \hat{M}(\yb)}{|\xb-\yb|} \ket{\psi_t}.
\end{equation}
For a single free particle of wave function $\psi$, it takes the perhaps more familiar form:
\begin{equation}
\frac{\upd}{\upd t} \psi(t,\xb) = \frac{i}{2\hbar m}\nabla^2\psi(t,\xb) + i\, \frac{Gm^2}{\hbar} \int \upd \yb\, \frac{|\psi(\yb,t)|^2}{|\xb-\yb|} \,\psi(t,\xb).
\end{equation}
The Schr\"odinger-Newton equation is a non-linear deterministic partial differential equation which brings serious conceptual difficulties if it is considered fundamental\footnote{Non-linear modifications of the Schr\"odinger equation are of course totally acceptable so long as they are effective, \eg approximately describing some large $N$ limit of a gas of particles.}. The problems arising from non-linear deterministic modifications of the Schr\"odinger equation have been widely discussed in the literature \cite{gisin1990,polchinski1991,bassi2015}. However, the assumptions made to derive the \emph{no-go} theorems are not always so clearly stated and it might be helpful to informally summarize what logically implies what.

\subsection{The trouble with non-linearity}

With non-linear terms in the Schr\"odinger equation, decoherence is no longer equivalent to collapse ``for all practical purposes''. As a result, what the Schr\"odinger-Newton equation says about the world (even only operationally) depends strongly on the interpretation of quantum mechanics one adopts, something that was noted already by Polchinski \cite{polchinski1991}. 

\begin{figure}
    \centering
    \includegraphics[width=0.6\textwidth]{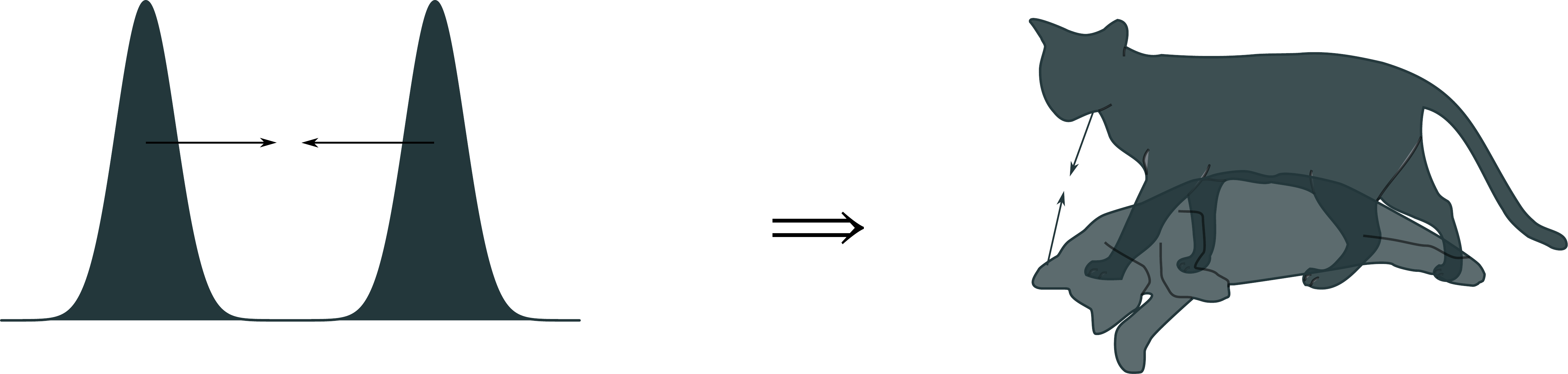}
    \caption{Without fundamental collapse, decoherence does not tame the SN gravitational interaction between different branches of the wave function. A single particle attracts itself but so does a (fully decohered) superposition of dead and live cat. (cat picture from \cite{doug})}
    \label{fig:cat}
\end{figure}

The first option is to consider an interpretation in which there is no fundamental collapse like in the Many-World interpretation or in the de Broglie-Bohm (dBB) theory. In that case the problem is that the non-linear gravitational term in the SN equation makes decohered branches of the wave function interact with each other. In the dBB picture, the ``empty'' part of the wave function gravitationally interacts with the non-empty one, thus influencing the motion of particles. In both cases, decoherence does not shield the dead and live cat from each other. The live cat constantly feels the attraction of its decaying ghost (see Fig. \ref{fig:cat}). Replacing cats by astrophysical bodies like the moon shows that the theory cannot be empirically adequate. The experiment carried out by Page and Geiker \cite{page1981} only makes this inadequacy manifest at the level of smaller macroscopic bodies\footnote{One sometimes hear that the Page and Geiker experiment is inconclusive because decoherence is not properly taken into account. This is a misunderstanding. The non-linear gravitational interaction in the SN equation is not tamed by decoherence and thus the conclusion about mean field couplings in \cite{page1981} is correct. The objection one could make to the Page and Geiker experiment is that there is no point in testing a fundamental theory on one system when it is already blatantly falsified on others.}.

\begin{figure}
    \centering
    \includegraphics[width=0.75\textwidth]{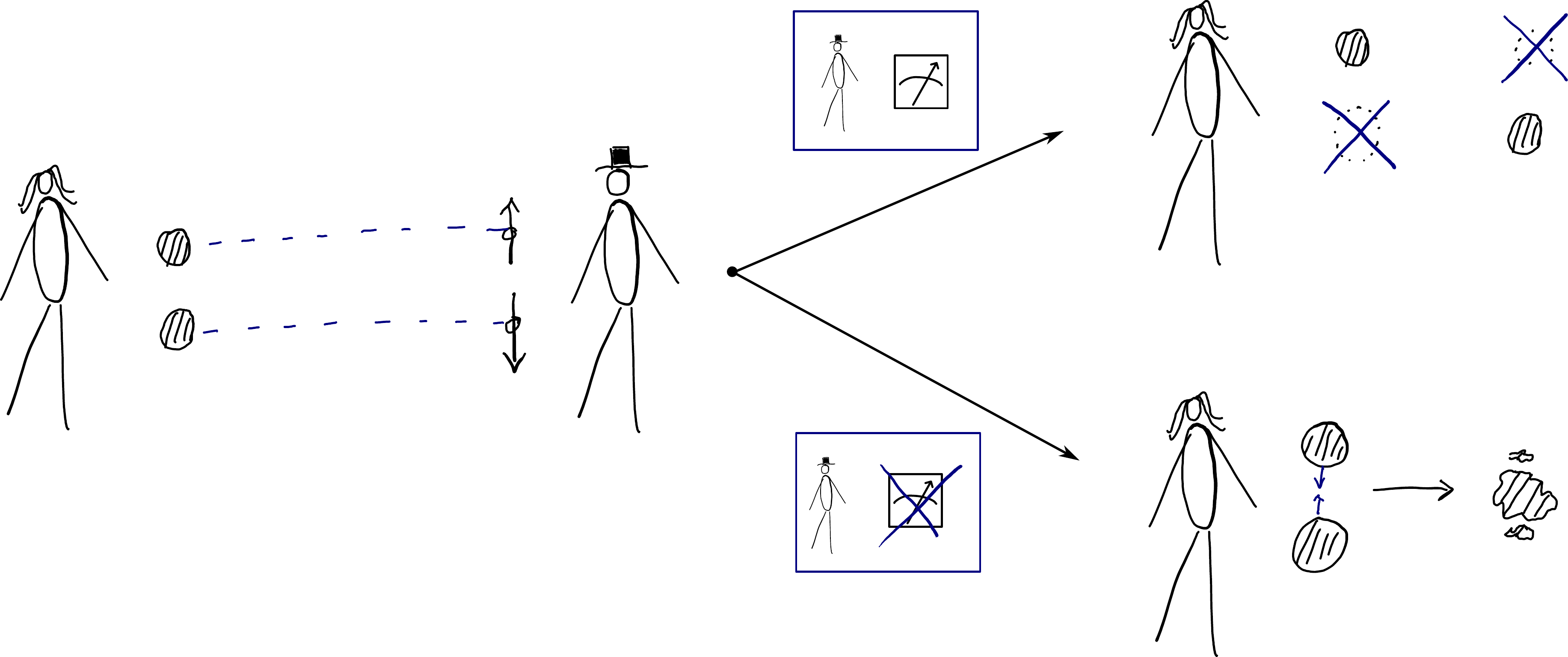}
    \caption{Alice and Bob start by sharing an EPR pair, which is encoded in a massive superposition on Alice's side. Bob can decide whether or not to measure his side. The subsequent evolution on Alice's side depends on this choice. This yields different probabilities to measure her mass \eg near the centroid of the initial superposition. This would allow Bob to communicate with Alice and even more seriously jeopardizes the validity of the Born rule. Notice that the gravitational force is not used directly as information carrier from Alice to Bob, but rather as a local non-linearity on Alice's side allowing to witness a wave-function collapse.}
    \label{fig:bob}
\end{figure}

The second option is to consider that a fundamental (and not effective) collapse of the wave function takes place in measurement situations. In that case, the difficulty is that the statistical interpretation of quantum states breaks down.
This can be seen quite straightforwardly. Let us consider a situation in which Alice and Bob share an EPR pair. On Alice's side, the spin degree of freedom is replaced by the position of a massive object $\ket{\mathsf{left}}$ and $\ket{\mathsf{right}}$. Let the initial state of the pair read:
\begin{equation}
\ket{\Psi}\propto \ket{\mathsf{left}}^{\rm Alice} \otimes \ket{\uparrow}^{\rm Bob} + \ket{\mathsf{right}}^{\rm Alice} \otimes \ket{\downarrow}^{\rm Bob}.
\end{equation}
I assume that Bob can choose to measure his spin at the beginning of the experiment. I write $\rho^{\rm Alice}_{\uparrow}$ (resp. $\rho^{\rm Alice}_{\downarrow}$) the density matrix of Alice if Bob measured $\uparrow$ (resp. $\downarrow$). I write $\rho^{\rm Alice}$ the density matrix of Alice if Bob does not measure his spin at all.
Because the Schr\"odinger-Newton equation is non-linear, it is easy to convince oneself that generically, after some time $t$:
\begin{equation}
\rho^{\rm Alice}(t)\neq \frac{\rho^{\rm Alice}_\uparrow (t)+ \rho^{\rm Alice}_\downarrow(t)}{2}.
\end{equation}
Intuitively, if Bob does measure his spin, Alice's wave function is collapsed in one position and nothing happens whereas if Bob does not measure his spin, Alice's wave function remains superposed and the two separated wave packets attract each other (see Fig. \ref{fig:bob}). Hence, the statistics on Alice's side crucially depend on what is done on Bob's side. At that point, the objection often raised is that this would allow instantaneous communication between Alice and Bob. The difficulty goes deeper. Even if one were to add delays or pick a preferred frame for collapse in order to avoid the paradoxes associated with faster than light signalling, a stronger objection would still remain. The essence of the difficulty, underlined by the previous example, is that reduced states cannot be trusted, \emph{even approximately}. To make predictions, Alice needs to know \emph{precisely} what Bob did and more generally everything that happened to whatever shared a common past with her. Reduced states (\ie typically the only objects one can consider in practice) have no statistical interpretation. Extracting predictions from the theory is a daunting if not impossible task.

To summarize, the Schr\"odinger-Newton equation (and typically all non-linear deterministic modifications of the Schr\"odinger equation) cannot be empirically adequate without an objective collapse mechanism. However, such a mechanism generically breaks the statistical interpretation of reduced states. This does not mean that it is strictly impossible to find an elaborate way to salvage this semiclassical approach and to one day provide an efficient way to compute predictions. That said, one can safely say that the Schr\"odinger-Newton equation, and thus the M\o ller-Rosenfeld approach to semiclassical gravity it derives from, is unappealing as a fundamental theory. In my opinion, trying to redeem it is hopeless and unnecessary: after all, the prescription of M\o ller and Rosenfeld (making expectation values gravitate) is completely arbitrary.

To my knowledge, all the conceptual difficulties encountered with the M\o ller-Rosenfeld approach to semiclassical gravity show up already in the Newtonian limit. That is, the problems of semiclassical gravity are quantum more than they are relativistic: naive hybrid theories, relativistic or not, are thought to have crippling problems. The bright side of this is that if we can construct consistent non-relativistic hybrid theories, we remove all the known serious objections against semiclassical gravity. 

\section{A consistent approach to Newtonian semiclassical gravity}\label{sec:4}

\subsection{Idea}
As the previous naive semiclassical gravity shows, using quantum expectation values as classical sources is a bad idea. But what kind of classical objects do we have at our disposal to consistently mediate an interaction between space-time and quantum matter? An option would be to consider a de Broglie-Bohm theory in which the gravitational field is sourced by the particles (or, in a relativistic context, by a Bohmian field). This was considered by Struyve \cite{struyve2015,struyve2017}. Although the theory is slightly less conceptually problematic --only the dead or the live cat source a gravitational field--, the statistical interpretation of states breaks down as before (``equivariance'' is lost). The approach of Derakhshani \cite{derakhshani2014}, built upon the model of Ghirardi Rimini and Weber \cite{ghirardi1986}, has essentially the same upside and downside\footnote{It should be noted that Derakhshani does mention the possibility of sourcing the gravitational field in a way that foreshadows the idea we shall present, although he does not explore it further in \cite{derakhshani2014}.}. It might be that the ultimate semiclassical theory has to destroy this appealing feature of quantum mechanics. However, in the absence of a statistical interpretation, it becomes astonishingly hard to use a physical theory to make reliable predictions, even in simple situations. It would thus be quite convenient if the statistical interpretation could survive. Fortunately, it can and the solution may surprisingly be guessed from the quantum formalism itself.

In the orthodox interpretation of quantum mechanics, there exists classical variables that both depend on the the quantum state and that can consistently back-react on it: measurement results. Indeed a measurement result depends, if only probabilistically, on the quantum state. Further, knowing a measurement result, an experimentalist can apply whatever unitary she likes on the quantum system without creating any kind of paradox. Experimentalists have been doing consistent semiclassical coupling for ages without knowing it! This arguably trivial discovery is what guided the construction of consistent hybrid semiclassical theories, dating back to 1998 and the work of Di\'osi and Halliwell \cite{diosi1998,halliwell1999}. These hybrid theories have been recently revived in the context of Newtonian gravity \cite{kafri2014,kafri2015,tilloy2016,tilloy2017,khosla2017,hall2017,tilloy2017grw}.

\begin{figure}
    \centering
    \includegraphics[width=0.45\textwidth]{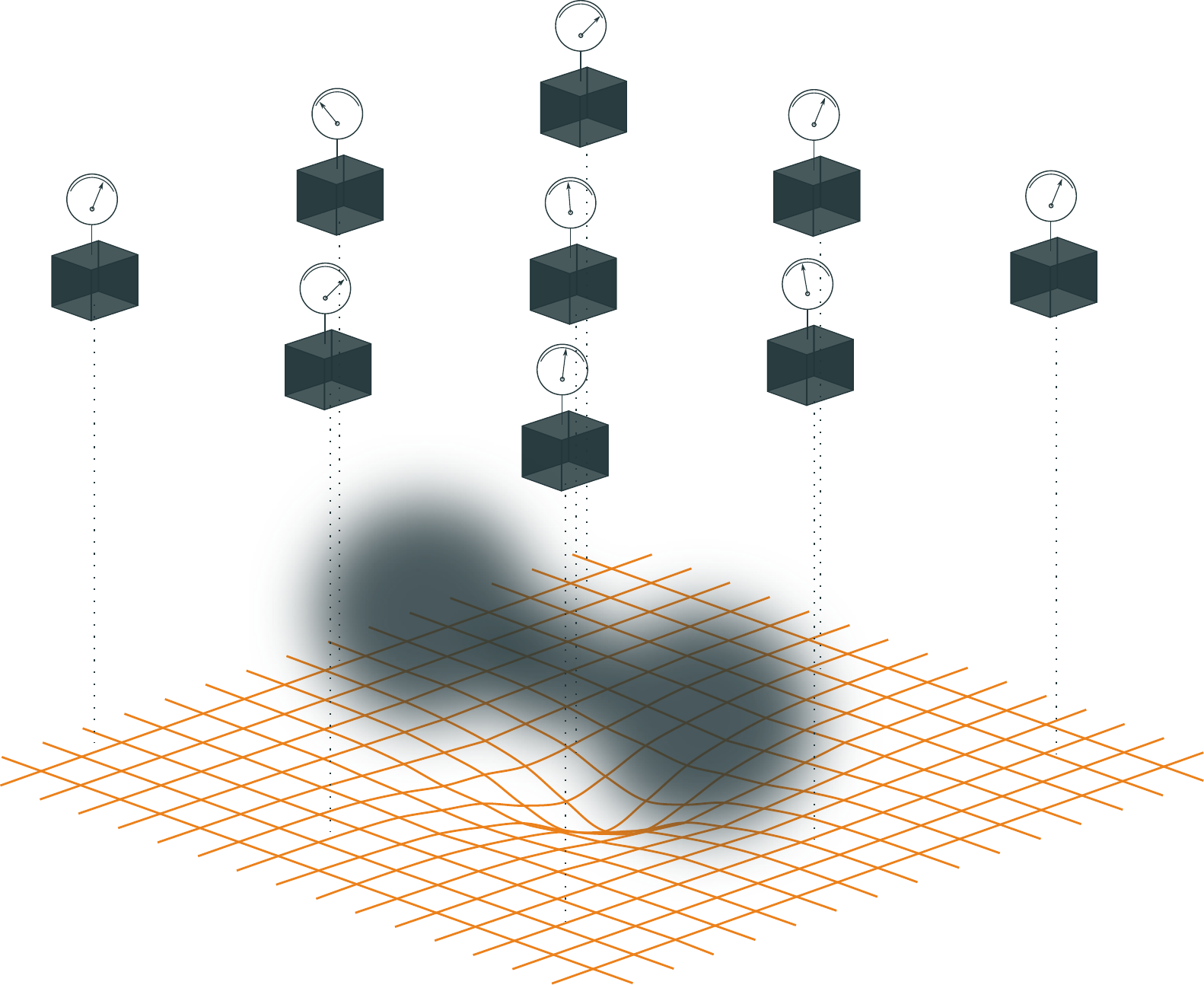}\hskip1.25cm \vline\hskip1.25cm
    \includegraphics[width=0.25\textwidth]{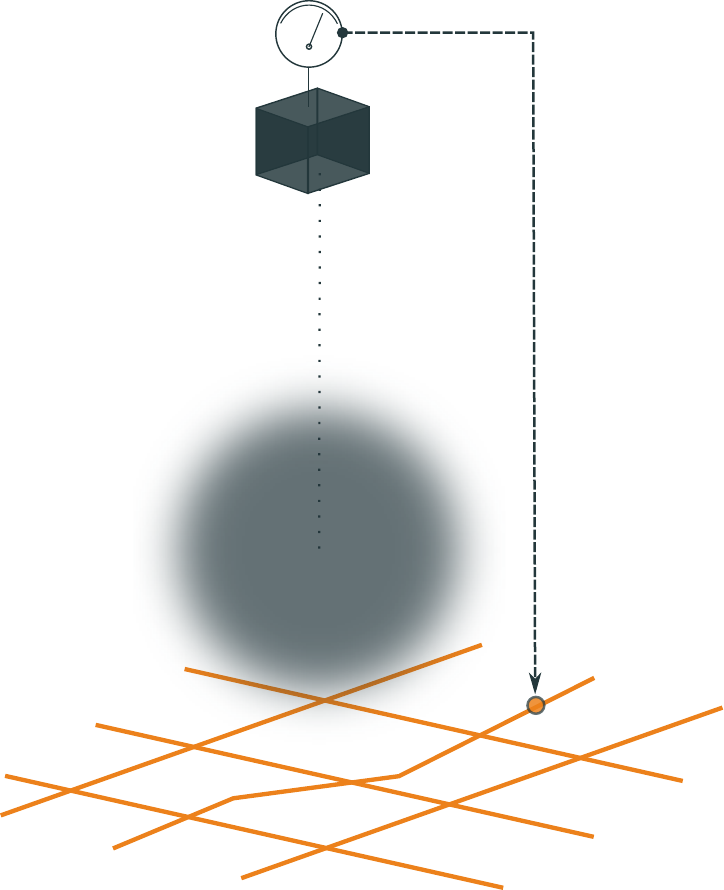}
    \caption{Illustration of the main idea -- The dynamics is \emph{as if}
 (left) space were covered with detectors weakly measuring the quasi local mass density and
 (right) the results associated to these measurement results were used to bend space-time, \ie in the Newtonian limit, source a gravitational field $\Phi$. At a \emph{formal} level, this is just measurement-based feedback.}
    \label{fig:detectors}
\end{figure}

I am not pushing here for the introduction of conscious observers or even any kind of physical measurement setup to couple quantum and classical variables. I am simply noting that at a purely mathematical level, the orthodox formalism already allows an interaction between quantum and classical variables in the form of a ``measurement and feedback'' scheme. One can use this fact as an intuition pump to construct a ``realist'' theory. We may just posit that the fundamental dynamics is \emph{as if} some mass density were weakly measured continuously and the associated measurement outcomes used as sources to curve space-time (see Fig. \ref{fig:detectors}).

\subsection{Measurement and feedback}\label{sec:meas}
Let me quickly recall how weak measurement and feedback work in a discrete time setup before going to the continuum. A quantum operations corresponding to a positive-operator valued measure (POVM) transforms the state in the following way:
\begin{equation}\label{eq:povm}
    \ket{\psi} \xrightarrow{\text{measurement}} \frac{\hat{N}_k \ket{\psi}}{\sqrt{\bra{\psi} \hat{N}_k^\dagger \hat{N}_k \ket{\psi}}},
\end{equation}
with probability $p_k=\bra{\psi} \hat{N}_k^\dagger N_k \ket{\psi}$. The label $k$ is the outcome of the measurement. The operators $\hat{N}_k$ verify as only constraint $\sum_k \hat{N}_k^\dagger \hat{N}_k = \mathds{1}$. Feeding back the result $k$ just amounts to apply a $k$-dependent unitary operator $\hat{U}_k$ on the state. That is, for a measurement immediately followed by a feedback, one has:
\begin{equation}
     \ket{\psi} \xrightarrow[+\; \text{feedback}]{\text{measurement}}  \hat{U}_k\frac{\hat{N}_k \ket{\psi}}{\sqrt{\bra{\psi} \hat{N}_k^\dagger \hat{N}_k \ket{\psi}}},
\end{equation}
with the same probability $p_k$. Notice that the full evolution is still a legitimate POVM with $\hat{N}_k\rightarrow \hat{B}_k= \hat{U}_k \hat{N}_k$. 

The label $k$ can be thought of as a classical variable that \emph{depends} on the quantum state (probabilistically) via the first measurement step and that acts on it via the second feedback step. Of course, unless one puts restrictions on the $\hat{N}_k$'s (\eg requiring mutually commuting self-adjoint operators), the decomposition into measurement and feedback is arbitrary and feedback can be included in measurement backaction. The picture in two step is however helpful as a guide to construct semiclassical theories.

With the discrete time setup in mind, one can already construct a consistent toy model of semiclassical gravity \cite{tilloy2017grw}. Let me promote the $k$ label to a continuous position index $\xb_f \in \mathds{R}^{3}$ and take $\hat{N}_k\rightarrow \hat{L}(\xb_f)$, acting on 1-particle wave-functions, where: 
\begin{equation}
   \hat{L}(\xb_f) =\frac{1}{(\pi r^2_C)^{3/4}} \mathrm{e}^{-(\hat{\xb}-\xb_f)^2/(2 r^2_C)}.
\end{equation}
Let the instants $t_f$ when the corresponding weak measurement is applied be randomly distributed according to a Poisson distribution of intensity $\lambda$. Between collapse events, we let the particle evolve according to the standard Schr\"odinger free evolution. The resulting state dynamics is that of the Ghirardi-Rimini-Weber (GRW) model \cite{ghirardi1986,bassi2013}, the simplest discrete collapse model. Naturally, the GRW model makes no reference to measurement but at a purely formal level, it is but the iteration of a simple POVM. This means that the collapse ``outcomes'' (or flashes) $(t_f,\xb_f)$ can be used to influence the subsequent evolution of the system. In our gravitational context, they are thus good candidates for a source of the gravitational field $\Phi$. We may thus propose:
\begin{equation}\label{eq:poissonflash}
    \nabla^2 \Phi(\xb,t)=\frac{4\pi  G m}{\lambda}\,  \delta^3(\xb-\xb_f)\, \delta(t-t_f),
\end{equation}
where $m$ is the mass of the particle and the factor $\lambda^{-1}$ is fixed to recover Newtonian gravity in the classical limit. Putting this gravitational field as an external source in the Schr\"odinger equation and inverting the Poisson equation \eqref{eq:poissonflash} simply gives that each discrete collapse is immediately followed by a unitary feedback:
\begin{equation}
    \hat{U}(\xb_f)=\exp\left(i \,\frac{G}{\lambda \hbar} \frac{m^2}{|\xb_f-\hat{\xb}|}\right).
\end{equation}
The empirical content of the model is contained in $\rho_t=\mathds{E}[\ket{\psi_t}\bra{\psi_t}]$, where $\mathds{E}[\,\cdot \,]$ denotes the average over measurement outcomes. It is easy to show that, as expected from our general argument, $\rho_t$ obeys a linear master equation which can be written explicitly:
\begin{equation}\label{eq:master2}
\partial_t \rho_t = -\frac{i}{\hbar}[\hat{H}_0,\rho_t] + \lambda \int \upd \xb_f \hat{B}(\xb_f)\rho_t \hat{B}^\dagger(\xb_f) \!- \!\rho_t,
\end{equation}
where $H_0$ is the Hamiltonian in the absence of collapse and $\hat{B}(\xb_f) = \hat{U}(\xb_f) \hat{L}(\xb_f)$.
Surprisingly, once straightforwardly extended to the many particle context, this rather brutal implementation of gravity does not (yet) clash in any obvious way with observations \cite{tilloy2017grw}. Indeed, one approximately recovers the expected Newtonian pair potential between a macroscopic source and test particles, the only situation that can so far be probed in experiments. Further, this implementation of gravity preserves the statistical interpretation of the quantum state by construction and is thus free of conceptual difficulty.

\subsection{Continuous setting}

The discreteness of the previous model may look aesthetically unappealing even if it does not make it empirically inadequate. With the help of continuous measurement theory, one may construct a smoother equivalent. Let me hint at how this can be done. 

The continuous measurement of an operator $\mathcal{O}$ (taken to be self-adjoint here for simplicity) is described by the following pair of equations \cite{jacobs2006,wiseman2009}:
\begin{align}
\partial_t \ket{\psi_t}&= \Big[-i H + \sqrt{\gamma}\left( \mathcal{O}-\langle\mathcal{O}\rangle_t\right) \eta_t - \frac{\sqrt{\gamma}}{2} \left( \mathcal{O}-\langle\mathcal{O}\rangle_t\right)^2 \Big] \ket{\psi_t} \label{eq:belavkin}\\
\mathscr{S}_{\hat{\mathcal{O}}}(t) &= \langle\mathcal{O}\rangle_t   + \frac{1}{2\sqrt{\gamma}}\,\eta_t \label{eq:signal}
\end{align}
with $\langle\mathcal{O}\rangle_t=\bra{\psi_t} \mathcal{O} \ket{\psi_t}$. The first line is the continuous equivalent of the discrete evolution \eqref{eq:povm} and $\mathscr{S}_{\hat{\mathcal{O}}}(t)$ is the associated ``signal''. The latter is just the continuous time counterpart of a measurement outcome. Its fluctuations around the quantum expectation value of $\mathcal{O}$ are proportional to $\eta_t$, a white noise process. Slightly non-rigorously, it is just a Gaussian process of zero mean and Dirac distribution as two point function $\mathds{E}[\eta_t\,\eta_s]=\delta(t-s)$. That is, one can obtain a signal $\mathscr{S}_{\hat{\mathcal{O}}}(t)$ with statistical average equal to the quantum expectation value of the desired operator at the price of some white noise $\eta_t$ and a non-trivial non-linear backaction on the quantum state. Notice however that one can trivially invert \eqref{eq:signal} and express $\eta_t$ as a function of $\mathscr{S}_{\hat{\mathcal{O}}}(t)$ to write \eqref{eq:belavkin} only as a function of the signal. The white noise process is introduced solely to make the probabilistic law of the signal explicit.

Equations \eqref{eq:belavkin} and \eqref{eq:signal} can be generalized to a continuous set of operators. Considering the simultaneous measurement of a smoothed mass density operator $\hat{M}_\sigma(\xb)$  in every point $\xb$ of space yields the dynamics of the Continuous Spontaneous Localization (CSL) model \cite{pearle1989,ghirardi1990,bassi2013}, the simplest continuous collapse model. Again, this latter model is usually defined without reference to measurement. Via this reformulation, I am only seeking a guide for theory building. The advantage of the measurement perspective is that it provides us with a natural candidate for a classical mass density field: the mass density signal $\mathscr{S}_{\hat{M}}(\xb)$. The tempting thing to do is to now make it source the gravitational field:
\begin{equation}\label{eq:continuouspoisson}
\nabla^2 \Phi(\xb) = 4\pi G \mathscr{S}_{\hat{M}}(\xb).
\end{equation}
As in \ref{sec:meas}, one may then carefuly\footnote{In this context, the slight vanity of using rigorous stochastic calculus is unfortunately a necessity because of the multiplicative noise term (understood in the It\^o convention) in \eqref{eq:belavkin}. Introducing a signal dependent potential in the Schr\"odinger equation requires some care. The potential has to act on the state infinitesimally after the measurement back-action. This can be done properly with It\^o calculus but the resulting stochastic differential equation is not trivial \cite{tilloy2016}.} add $\Phi$ as an external source in the Schr\"odinger equation and invert the Poisson equation \eqref{eq:continuouspoisson} to get a perfectly consistent evolution for quantum states. I shall not go further but mention that this is one of the models that was introduced in \cite{tilloy2016} and that it is again a priori compatible with observations.

Actually, the continuous case allows for more refinements and there are infinitely many ways (parameterized by a positive real function) in which one can measure the mass density operator in every point of space. Heuristically, this comes from the freedom in correlating the signals of detectors in different regions of space. The general case was nonetheless treated as well in \cite{tilloy2016}, yielding many candidates for Newtonian semiclassical gravity. It was further shown in \cite{tilloy2017} that requiring a minimal amount of decoherence allowed to pin down a unique model (which happens to have the decoherence of the Di\'osi-Penrose model).

\subsection{Difficulties}\label{sec:difficulties}  

The technicalities of continuous measurement theory and It\^o calculus put aside, the continuous case works very much like the discrete. In both cases, whether one assumes gravity acts by instantaneous kicks or more smoothly, one obtains simple semiclassical models that are so far compatible with what is known about gravity in the Newtonian limit. That said, it is important to fairly present the difficulties and unappealing features that come with this idea.

The main drawback of our approach is that it introduces a short distance regularization of the gravitational potential. To create the mass density source of the gravitational field, I have formally relied on the measurement (continuous in time or discrete) of a smeared version of the mass density operator. This was necessary to make measurement back-action finite and avoid infinite positional decoherence. This regularization of the source propagates to the gravitational force and the $1/r$ pair-potential of gravity thus breaks down at short distances. The cutoff cannot be too small (smaller than the nuclei radius) otherwise the models I have briefly introduced are already falsified. This regularization is perhaps the most ad hoc and unappealing aspect of these models. The only comfort one may keep is that the regularization scale can be experimentally constrained on both ends, at least in principle: small cutoff length can be falsified by measuring excess positional decoherence and large cutoff length can be falsified by measuring deviations from the $1/r^2$ law of gravity \cite{tilloy2017grw,tilloy2017}.

The second obvious limitation of the approach I have reviewed is the lack of a straightforward Lorentz invariant generalization. The main reason is that it is extremely difficult to construct simple relativistic continuous measurement models. In particular, the space smearing necessary to make decoherence finite in the non-relativistic case seems to require some smearing in time in the relativistic regime. To my knowledge, there exists no continuous measurement model in which the necessary non-Markovianity this implies can be implemented in a sufficiently explicit way. There are nonetheless a number of promising options currently explored, either dropping the measurement interpretation \cite{tilloy2017qft} or going to General Relativity directly to make use of its additional structure \cite{bedingham2016,juarezaubry2017}.

Even if one had a fully consistent theory of relativistic continuous measurement, one could not straightforwardly generalize the approach by replacing the Poisson equation by Einstein equation and the mass density by the stress-energy tensor. Indeed, the analog of the ``signal'' in this relativistic context will typically not be conserved, something that is forbidden by Einstein equations. There are proposals to deal with non-conserved stress energy tensors using unimodular gravity \cite{josset2017} and they yield interesting predictions especially regarding the cosmological constant. This would however add one further subtlety making the whole story possibly less endearing. 

Finally, after enumerating the difficulties that are still in the way, we should not forget what the objective was. My aim was to show that the arguments used against semiclassical gravity (and that hold in the Newtonian limit already) can easily be bypassed. The way I have done so may not be straightforward to generalize nor aesthetically pleasing, but it achieves its purpose: providing a counterexample. The standard objections against semiclassical gravity do not hold.

\section{Discussion}\label{sec:5}

\subsection{Space-time and the measurement problem}

The program I have presented is interesting in that it ties together semiclassical gravity and the measurement problem. To create a gravitational field, one uses a classical source formally obtained from a (discrete or continuous) measurement setup. Once this formalism is taken as fundamental and before gravity is turned on, the model is simply a vanilla collapse model. Such models have been introduced more than 30 years ago as an ad hoc but working solution to the measurement problem.  Interestingly, a compact way to understand how collapse models dissolve the measurement problem is to introduce an explicit primitive ontology, that is a candidate for the ``stuff'' that makes the fabric of reality \cite{allori2014,allori2015}. What I have done is simply to source gravity with one standard candidate for a primitive ontology (the flashes in the discrete and their more confidential field counterpart in the continuum).
Hence, starting from the need to have classical sources, one is naturally guided towards a class of models that fix quantum theory. One needs to pay the price of collapse models only once, either to help with the measurement problem or to help with quantum classical coupling; the other comes for free.

One may speculate that this connection between the problems of measurement and of semiclassical gravity extends beyond the specific approach used to derive it. A crude (and unfortunately very non-operational) definition of semiclassical gravity would be that it is a theory in which gravity is mediated by a classical field. The existence of such a field without superposition necessarily dissolves the measurement problem: there is now something out there, not superposed, that can be used to specify measurement results. The problem then becomes one of empirical adequacy: can we understand our world and experiences making reference only to functions of this space-time? Provided the classical metric retains some properties it would have in classical general relativity, a candidate for a primitive ontology would be the stress-energy tensor distribution reconstructed from Einstein's equations. Presumably, one could understand measurement outcomes from a simple coarse graining of this distribution of matter. Semiclassical gravity does seem to go hand in hand with a down to earth understanding of quantum theory.

\subsection{Comparing the alternative}

With semiclassical gravity, space-time is as simple as it is in general relativity, no more no less. Quantum mechanics, once confined to matter, requires no paradigm shift. This compares favorably with what happens if one requires that gravity itself be quantized. In that case, depending on the approach taken, one needs to tell a far subtler story of classical and quantum emergence.

In string theory and causal set theory for example, there is no 4 dimensional space-time at a fundamental level, even classically. The first thing one needs to explain is how a 4 dimensional space-time emerges at large scales from the compactification of a higher dimensional manifold or from the coarse graining of a branching causal structure. This story is not entirely trivial but one should not forget that it only the first easiest step. One then needs to understand the emergence of a classical space-time from the effective dynamics of some wave-functional on 3-metrics, a problem at least as hard as the measurement problem. There is no reason to believe that it is impossible to do so, but the story will inevitably be quite elaborate. If the alternative straightforward semiclassical story is not experimentally falsified nor theoretically forbidden, what ground do we have to take complicated explanations of emergence seriously? 

Experiments will ultimately decide if the full quantization route really has to be taken. In the meantime, it is certainly worth exploring whether or not one can understand, even only in principle, the subtle emergence of space-time from a quantized theory of gravity. Yet one should probably keep as the default option (or null hypothesis) the simpler semiclassical story where space-time is just there and not emergent. As mentioned in \ref{sec:difficulties}, the semiclassical route is not without open questions but its metaphysical clarity definitely calls for further exploration. It is pressing to know how far it can be pushed, if only to eventually reach the conclusion that quantizing gravity is really necessary.

\subsection{Resisting the straw man}
Constructing working theories of quantum gravity is a tough and ungrateful task. One risks working for years on an approach that does not even remotely describes the physical world. For that reason, no-go theorems --excluding the paths not taken-- provide a warm comfort. Perhaps the route is long, but at least we go in the only possible direction. But no-go theorems, whether they are real theorems or just heuristics, are rarely as strong as they seem, especially after their message has been inflated by the folklore in the field. As I hope to have shown, no-go arguments should be treated with caution especially in the context of gravity where they are especially comforting.


One should not be too quick to bury semiclassical gravity if a specific approach turns out to be falsified. We should make sure that we do not disprove straw men to justify grand claims, and stick with the boring option for as long as it can be reasonably\footnote{All the difficulty is of course in knowing when this defense becomes unreasonable.} defended.

\section{Summary}
It is possible to construct working toy-models of semiclassical gravity. These models should perhaps not bee taken too seriously as they are explicit only in the Newtonian limit. However, they provide a counter example to the non-relativistic arguments usually invoked to disregard theories with a tangible, non-superposed space-time. There is thus no definitive reason, theoretical or experimental, cornering us to a quantized space-time. The subtle emergence of our intuitive space-time from more complicated quantum versions may not have to be unraveled. 

Two things could change this state of affairs. The quantization of the gravitational field could show itself in experiments. We would then have no choice but to try and understand space-time as an emergent notion. Alternatively, one could find an interpretation of quantized gravity far ``simpler'' than the apparently straightforward semiclassical story we have presented. This would be no simple task, especially as the semiclassical route seems to also  brings with it a natural solution to the measurement problem. Given that this challenge has not been met yet and given that experiments are still silent about the true nature of gravity, it is tempting to stick with our good old classical space-time.

\paragraph*{Acknowledgments:}
I thank Lajos Di\'osi for helping shape my views on this subject. I am grateful to an anonymous referee for sharp comments on an earlier version of this manuscript, submitted as an essay to the \textit{Beyond Spacetime} 2017 contest. This work was made possible by support from the Alexander von Humboldt foundation and the Agence Nationale de la Recherche (ANR) contract ANR-14-CE25-0003-01.

\bibliographystyle{apsrev4-1}
\bibliography{main}

\end{document}